\documentclass[11pt]{article}

\usepackage{graphicx}
\usepackage{slashed}
\newcommand {\ga} {\ {\raise-.5ex\hbox{$\buildrel>\over\sim$}}\ }
\newcommand {\la} {\ {\raise-.5ex\hbox{$\buildrel<\over\sim$}}\ } 
\begin{document}

\title{Cogenerating and Pre-annihilating Dark Matter by a New Gauge Interaction
in a Unified Model}

\author{{\bf S.M. Barr} \\
Department of Physics and Astronomy \\ 
and \\
Bartol Research Institute \\ University of Delaware \\
Newark, Delaware 19716 \\
\\
{\bf Robert J. Scherrer} \\
Department of Physics and Astronomy \\ 
Vanderbilt University \\
Nashville, Tennessee 37235}

\maketitle

\begin{abstract}
Grand unified theories based on large groups (with rank $\geq 6$) are a natural context for dark matter models. They contain Standard-Model-singlet fermions that could be dark matter candidates, and can contain new non-abelian interactions whose sphalerons convert  baryons,  leptons, and dark matter into each other, ``cogenerating" a dark matter asymmetry comparable to the baryon asymmetry. In this paper it is shown that the same non-abelian interactions can ``pre-annihilate" the symmetric component of heavy dark matter particles $\chi$, which then decay late into light stable dark matter particles $\zeta$ that inherit their asymmetry. We derive cosmological constraints on the parameters of such models. The mass of $\chi$ must be $< 3000$ TeV and their decays must happen when $2 \times 10^{-7} < T_{dec}/m_{\chi} < 10^{-4}$. It is shown that such decays can come from $d=5$ operators with coefficients of order $1/M_{GUT}$ or $1/M_{P\ell}$.  We present a simple realization of our model based on the group $SU(7)$.
\end{abstract}

\section{Introduction}

In recent papers it has been argued that grand unification based on groups larger than $SU(5)$ has several features that make it a natural context for theories of dark matter \cite{unified-DM, BC-sphaleron}. 
(1) The fermion multiplets of such unified groups typically contain Standard Model singlets, which could play the role of dark matter. (2) Unified theories can have accidental global symmetries (a well-known example is $B-L$ in minimal $SU(5)$), and one of these could be the quantum number $X$ that stabilizes dark matter. And (3), unification groups of rank 6 or larger can contain additional non-abelian interactions, which can play a role in generating dark matter. One way this can happen is through the sphaleron processes \cite{sphaleron} of the new non-abelian interaction, which can convert baryons and leptons into dark matter particles, thus giving a scenario in which dark matter and baryonic matter are ``co-generated." 

Many papers have explored the idea that sphalerons are involved in the genesis of dark matter. Usually, these scenarios involve only the sphalerons of the electroweak interactions \cite{BCF, ADM-sphaleron}. However, as emphasized in \cite{ADM-nonSMsph} (and later in \cite{BC-sphaleron}), the idea that the sphaleron processes of a new non-Standard Model interaction do the co-generation has several attractive features. First, if electroweak sphalerons produced the dark matter asymmetry, the dark matter particles have to be chiral under electroweak $SU(2)$, which makes them more visible both directly and through their loop effects, severely constraining such models. By contrast, to be produced by the sphalerons of a new interaction, the dark matter particles need only be chiral under the new interaction and can be vectorlike under the electroweak $SU(2)$, which allows them to be heavy and to have negligible loop effects on Standard Model predictions. Such models are therefore less severely constrained and much easier to build. Second, having two kinds of sphalerons --- those of the electroweak $SU(2)$ and those of a new gauge interaction --- means that there are {\it two} sphaleron equilibrium conditions, which is enough to determine the ratios of $B$, $L$, and $X$. That means (under the assumption that the dark matter is almost purely asymmetric, i.e. $n_{\overline{DM}} \ll n_{DM}$) that the number density of dark matter particles today can be calculated, giving a definite (though of course model-dependent) prediction for the mass of the dark matter particle. As sphaleron processes typically make $B$, $L$, and $X$ comparable, and because $\Omega_B$ and $\Omega_{DM}$ are comparable, one typically finds that the mass of the dark mass particle is GeV-scale . 

To have the dark matter be asymmetric \cite{nussinov, BCF, ADM-sphaleron, ADM-nonSMsph, ADM-other, Graesser, LYZ}, there must have been some processes in the early universe that annihilated the ``symmetric component" of the dark matter. That is, most of the anti-dark matter particles must have annihilated leaving just the asymmetry (as was the case with baryons). This is a non-trivial issue if the dark matter particles have masses of order a GeV, as then
annihilation via electroweak interactions is not efficient enough and would have left a relic abundance of anti-dark matter particles that was far too large. One way to solve this problem \cite{BC-sphaleron} is to posit the existence of a light field (typically it must have mass less than the weak scale) that mediates an interaction {\it by which} dark matter particles and their anti-particles can annihilate, as well as some new massless (or extremely light) particles {\it into which} they can annihilate. Thus the existence of an entire sector of light particles, besides the dark matter particles themselves, would be postulated {\it ad hoc}. 

In this paper, we point out another way: The same new non-abelian gauge interactions whose sphalerons co-generate dark matter in the first place, are also able to annihilate the symmetric component of the dark matter. This leads to a very economical, tightly integrated scenario, in which everything comes from unification in larger groups: the dark matter fermions themselves, and the new interaction that generates them and annihilates the symmetric component. 

At first, it might seem that the new non-abelian gauge interactions whose sphaleron processes generate the dark matter asymmetry cannot do the annihilation efficiently enough, any more than the weak interactions could, for the scale characterizing their breaking must be even larger than the weak scale. However, as we show in this paper, they can do it through a mechanism that we call``pre-annihilation." This mechanism was proposed in the context of other scenarios by Kitano and Low \cite{KitanoLow} and by Kang and Li \cite{KangLi}. (It can also be seen as a limiting case of the ``exodus" mechanism of Unwin \cite{Unwin}.) Because the scenario we are describing is tightly integrated and to a large extent determined by the grand unified structure, one can use cosmological observables to constrain it. 

The idea of pre-annihilation is that an asymmetry in $X$ is first produced among some heavy metastable dark particles, which are able to annihilate efficiently enough to make negligible the number of their anti-particles. These heavy dark particles then decay late, out of equilibrium, to stable dark particles that ``inherit" their asymmetry and form the dark matter that is observed today. (We shall call the heavy dark particles $\chi$ and the light stable dark particles $\zeta$.) We shall show that the lifetime of these late decays must fall within a certain window. A nice feature of the scenario we are discussing is that these decays can result from effective dim-5 operators that would give the right lifetime if they were suppressed by $O(1/M_{P\ell})$ or $O(1/M_{GUT})$. 

The basic cosmological scenario we shall discuss has the following stages. When the temperature is above the mass scale
of the heavy dark particles, $m_{\chi}$ (which is much larger than the weak scale but less than $3 \times 10^6$ GeV), the
sphaleron processes of the new non-Standard Model gauge interaction, whose gauge group we shall call $G_*$, produce an
$X$ asymmetry comparable to the baryon asymmetry. This asymmetry is in the form of an excess of $\chi$ over their
anti-particles $\chi^c$. As the temperature drops below $m_{\chi}$, the $\chi$ and $\chi^c$ annihilate by means of the
$G_*$ gauge interaction, leaving only the excess $\chi$ due to the $X$ asymmetry. At a later epoch (when $T$ is between
$10^{-4} m_{\chi}$ and $2 \times 10^{-7} m_{\chi}$), each heavy dark particle $\chi$ decays via interactions suppressed by $1/M_{GUT}$ or $1/M_{P \ell}$ into one light dark matter particle $\zeta$ plus some Standard Model quarks and leptons that equilibrate. The $\zeta$ are singlets under all gauge interactions --- except possibly ones broken at $M_{GUT}$ --- and are never in thermal equilibrium. (At some very early time, their density was driven to exponentially small values by inflation, and the only such particles present today are those caused by the decay of the heavy dark particles.) Thus the stable light dark matter particles inherit the asymmetry produced at very large scales. Because they have never been in equilibrium the dark matter particles are produced with a mean momentum significantly higher than the photon temperature. Depending on the parameters of the model, this scenario can give either cold or warm dark matter.

As noted above, some of the ingredients of the scenario we are proposing and studying here
have been considered elsewhere, such as the use of non-Standard Model sphalerons to co-generate dark matter \cite{ADM-nonSMsph, BC-sphaleron}, and the pre-annihilation of the symmetric component of dark matter \cite{KitanoLow,KangLi,Unwin}. What is new here is that we combine these elements into a relatively simple scenario and show that it emerges naturally in the context of grand unification based on groups of rank 6 or greater. And within models of the type we are proposing, we derive significant new limits coming from several considerations. These include (a) the condition that the decays of the heavy unstable dark matter not be so slow as to make the resulting stable dark matter hot, and (b) that the decays of the heavy unstable dark matter not be so rapid as to regenerate a large symmetric component of dark matter. These limits give a relatively narrow window of decay lifetimes of the heavy dark matter, which we show turns out to be nicely consistent with decay via GUT- or Planck-suppressed operators that one might expect to exist in realistic models.

All of this we illustrate in a specific grand unified model based on $SU(7)$. In that model, we show how the various critical ingredients tend naturally to exist in grand unified models with large gauge groups. The model illustrates, in particular, how the dark matter and ordinary matter particles are unified in GUT multiplets; how the $G_*$ interactions are unified with the Standard Model interactions; how in a typical case there are quarks, leptons and dark matter particles that transform under $G_*$, as needed for the co-generation mechanism; and how the quantum number $X$ can arise as an accidental symmetry in a grand unified model. 
 
We shall first discuss the cosmology of the scenario in more detail in section 2, and then the details of the illustrative $SU(7)$ model in section 3. 

\section{The details of pre-annihilation}
 
The new gauge interaction based on the group $G_*$ is broken at a scale $M_*$ by the VEV of a Higgs field that we shall call $\Omega$. It is the sphalerons of the $G_*$ interaction that will violate $X$ and relate the $X$ asymmetry to the $B$ and $L$ asymmetries. It is also the $G_*$ gauge interaction that will mediate the annihilation of the heavy dark particles ($\chi$) with their anti-particles ($\chi^c$). The heavy dark particles ($\chi$) which have $X\neq 0$, are neutral under the Standard Model gauge interactions, but transform in a chiral way under $G_*$. (It is because of this chirality, of course, that the $G_*$ sphalerons violate $X$.) The particles $\chi$ and $\chi^c$ obtain masses of order $M_*$ from the VEV of $\langle \Omega \rangle$. We can therefore denote the scale of the breaking of $G_*$ by $m_{\chi}$ (instead of $M_*$) and shall do so from now on. Some of the Standard Model quarks and leptons will also transform under $G_*$, so that $G_*$ sphalerons also violate $B$ and $L$. (A simple model having all these features will be presented in section 3.)

The light dark particles ($\zeta$) and anti-particles ($\zeta^c$) are singlets under $SU(3)_c \times SU(2)_L \times U(1)_Y \times G_*$. They couple to the other particles of the model through dimension-5 operators suppressed by $1/M_{GUT}$ or $1/M_{P \ell}$. 

We assume that at very early times inflation wiped out $B$, $L$, and $X$. After
(or during) reheating an asymmetry of one or more of these quantum numbers was generated.
It will not matter for our discussion which asymmetry was generated or what process generated it. We assume that no significant number of $\zeta$ and $\zeta^c$ particles appeared during reheating, because they have no coupling to the inflaton and only couple to other light particles through GUT-scale-suppressed operators. (This will put an upper limit on the reheat temperature that depends on the GUT scale, as will be discussed later.)

When $T$ was large compared to $m_{\chi}$, the $G_*$ sphaleron processes were happening and produced an equilibrium among $B$, $X$, and $L$, making these  
asymmetries comparable.  (Of course, as is common, we use the term``sphaleron process" loosely to mean any anomalous $G_*$ interaction, not just the ones involving tunneling via the sphaleron configuration.) Let us define $\epsilon_B \equiv (n_B - n_{\overline{B}})/s$, $\epsilon_L \equiv (n_L - n_{\overline{L}})/s$, and $\epsilon_X \equiv (n_X - n_{\overline{X}})/s$. In order to have $\Omega_{DM} \approx 5 \Omega_B$, one requires that $\epsilon_X \approx (5 m_p/m_{\zeta}) \epsilon_B$, where $m_p$ is the proton mass.

When $T$ dropped below $m_{\chi}$, the number of
$\chi$ and $\chi^c$ became Boltzmann suppressed as they annihilated into lighter states.
These annihilations would have happened primarily through $G_*$ gauge interactions (and partly through $\Omega$ exchange). (Therefore,
they would not have annihilated into the light dark particles $\zeta + \zeta^c$.)   The relic
abundance of an asymmetric particle can be expressed in terms of the asymmetry parameter $r$,
which is defined to be
\begin{equation}
r = \frac{n_\chi^c}{n_\chi}.
\end{equation}
For an annihilation cross section scaling as $\sigma_0 (m_{\chi}/T)^{-n}$,
the value of $r$ at late times, $r_\infty$, is given by \cite{Graesser,LYZ} (see also earlier calculations in Refs. \cite{Earlierfreezeout}):
\begin{equation}
\label{rinf1}
r_\infty \approx \exp\left(-\frac{\epsilon_\chi \lambda \sqrt{g_{*f}}}{x_f^{n+1}(n+1)} \right),
\end{equation}
where $\lambda = 0.264~ M_{P\ell} m_{\chi} \sigma_0$ (with $M_{P\ell}$ denoting the Planck mass), $g_{*f}$ is the effective
number of relativistic degrees of freedom when the annihilations freeze out at temperature $T_f$, and $x_f = m_{\chi}/T_f$.
(We will use $g_*$ throughout to denote relativistic degrees of freedom, with the subscript labeled
to denote the temperature at which this quantity is evaluated in each case).  Taking the annihilation
in our model to be $s$-wave ($n=0$) with $\sigma_0 = \alpha_*/m_{\chi}^2$, Eq. (\ref{rinf1}) simplifies to
\begin{equation}
\label{rinf2}
r_\infty \approx \exp\left(- \frac{0.264~\epsilon_\chi \alpha_* M_{P\ell} \sqrt{g_{*f}}}{m_{\chi} x_f} \right)
\end{equation}

Our goal is to achieve $\epsilon_\chi \sim 10^{-10}$ (the same order of magnitude as for baryons).
Somewhat arbitrarily, we will take the dividing line between symmetric ($r_\infty = 1$) and
asymmetric ($r_\infty \ll 1$) freeze-out to be the point at which the argument in the exponential of Eq. (\ref{rinf2}) is $-1$;
asymmetric freeze-out will occur when the absolute value of this argument is $> 1$.
For the model parameters considered here,
$\sqrt{g_{*f}} \sim 20$ and $x_f \sim 20$. (As noted in Ref. \cite{Graesser}, $x_f$ is little changed from its symmetric
value unless $r_\infty \ll 1$).  Assuming that $\alpha_* \sim 10^{-2}$, we find that the $\chi$ freeze-out will be asymmetric
as long as $m_{\chi} < 3 \times 10^6$ GeV = 3000 TeV.  When this limit is satisfied, to a good approximation the only heavy dark particles
$\chi$ that remain after freeze-out are those that existed due to the $X$ asymmetry.

As mentioned, we assume each $\chi$ decays
via a dimension-5 GUT suppressed operator into a light dark particle $\zeta$ plus  
Standard Model particles. Let us call the coefficient of this dimension-5 operator 
$c_5/M_{GUT}$, where $M_{GUT}$ is the scale where the Standard Model gauge couplings unify, and $c_5$ is a dimensionless number. The $\chi$ decays will happen when the temperature of the universe is $T_{dec}$, given by

\begin{equation}
\left( \frac{c_5}{M_{GUT}} \right)^2 m_{\chi}^3 \sim \sqrt{g_{*dec}} \frac{T_{dec}^2}{M_{P\ell}}
\;\; \Longrightarrow \;\; \frac{T_{dec}}{m_{\chi}} \sim g_{*dec}^{-1/4} \frac{\sqrt{m_{\chi} M_{P \ell}}}{(M_{GUT}/c_5)}.
\end{equation}

So far, we have been giving a fairly standard analysis of asymmetric freeze out. We now show that there are two considerations that yield 
both an upper bound and a lower bound on $T_{dec}$ that must be satisfied to achieve a satisfactory asymmetric dark matter scenario. The upper bound comes from considering the $\chi^c$ that decay into $\zeta^c$ when $T \sim m_{\chi}$, when the 
$\chi^c$ are not Boltzmann suppressed. Of course, only a small fraction do, since
these decays have a lifetime much larger than the age of the universe when $T \sim m_{\chi}$.
But this fraction must be smaller than $\epsilon_X \cong \frac{ 5 m_p}{m_{\zeta}} \epsilon_B \sim 10^{-10}$ if one is not to produce
a number of $\zeta^c$ that is comparable to the $\zeta$ asymmetry. This gives a lower bound on the lifetime of the decay, and thus an upper bound on $T_{dec}$.
One has

\begin{equation}
\left( \frac{n_{\zeta^c}}{s} \right)_{T_{dec}} \sim 
\frac{t_{m_{\chi}}}{t_{dec}} \left( \frac{n_{\chi^c}}{s} \right)_{T \sim m_{\chi}}
\sim \frac{t_{m_{\chi}}}{t_{dec}} \frac{1}{g_{*{m_{\chi}}}},
\end{equation}

\noindent where $t_{m_{\chi}}$ is the age of the universe when $T \sim m_{\chi}$,
and $t_{dec}$ is the age of the universe when $T = T_{dec}$. Demanding that this be less than 
$\epsilon_X \cong \frac{ 5 m_p}{m_{\zeta}} \epsilon_B$, gives

\begin{equation}
\frac{t_{m_{\chi}}}{t_{dec}} \frac{1}{g_{*{m_{\chi}}}} =
\frac{(\sqrt{g_{*{m_{\chi}}}} m_{\chi}^2/M_{P \ell})^{-1}}{(\sqrt{g_{*dec}} T_{dec}^2/M_{P \ell})^{-1}} \frac{1}{g_{*{m_{\chi}}}} < \frac{ 5 m_p}{m_{\zeta}} \epsilon_B
\;\; \Longrightarrow  \frac{T_{dec}}{m_{\chi}} < \left( \frac{5 m_p}{m_{\zeta}} \right)^{1/2}
\frac{g_{*{m_{\chi}}}^{3/4}}{g_{*dec}^{1/4}} \epsilon_B^{1/2} \sim 10^{-4} .
\end{equation}
We have neglected two additional effects that will tend to suppress the decays of the $\chi^c$.
At $T \sim m_{\chi}$, the decay rate will be Lorentz suppressed, and inverse decays
will be energetically allowed.
Both of these will decrease the number of decay-produced $\zeta^c$, so our limit is a conservative one.

The lower bound on $T_{dec}$ comes from requiring that the dark matter streaming length be small enough to allow for the formation of the observed
large-scale structure. When the dark matter particles $\zeta$ are produced by decays of the
$\chi$, they have a momentum that is of order $m_{\chi}/3$ (assuming three-body decays).  This momentum redshifts as the inverse of the scale factor, so
a lower $T_{dec}$ allows for less redshift prior to the epoch of structure formation, resulting in a larger velocity and
corresponding larger
free-streaming
length.

The comoving free-streaming length $\lambda_{FS}$ up to the time $t_{eq}$ when the matter and radiation
densities become equal and density perturbations can begin to grow, is given by \cite{KolbTurner}
\begin{equation}
\label{lambdaFS}
\lambda_{FS} = \int_0^{t_{eq}} v(t^\prime) \frac{R_0}{R(t^\prime)} dt^\prime,
\end{equation}
where $R$ is the scale factor, and the zero subscript refers to present-day quantities.
Using the approximation that the velocity is $v=1$ when $\zeta$ is relativistic and scales as
$1/R$ when it becomes nonrelativistic at $t_{NR}$, we obtain \cite{KolbTurner}
\begin{equation}
\lambda_{FS} \approx t_{NR}\left(\frac{R_0}{R_{NR}}\right)\left[2+ \ln\left(\frac{t_{eq}}{t_{NR}}\right)\right]. 
\end{equation}

The momentum of $\zeta$ at scale factor $R$ is given by
$p = (m_{\chi}/3)(R_{dec}/R)$, and $\zeta$ becomes nonrelativistic when $p \approx m_\zeta$, so we have
\begin{equation}
T_{NR} = T_{dec} \left(\frac{3 m_\zeta}{m_{\chi}}\right)\left(\frac{g_{*dec}}{g_{*NR}}\right)^{1/3},
\end{equation}
where we have used the fact that
$g_* (RT)^3$ is constant \cite{KolbTurner}.  In the radiation-dominated era, $t = 0.30 g_*^{-1/2} M_{P\ell} T^{-2}$\cite{KolbTurner},
so we obtain
\begin{equation}
t_{NR} = 0.03 M_{P\ell}\left(\frac{m_{\chi}}{m_\zeta}\right)^2 T_{dec}^{-2}~ g_{*NR}^{1/6} ~g_{*dec}^{-2/3}.
\end{equation}
Further, $R_0/R_{NR}$ in Eq. (\ref{lambdaFS}) can be written in terms of $T_{dec}$:
\begin{equation}
\frac{R_0}{R_{NR}}
= \frac{3 m_\zeta}{m_{\chi}}\left(\frac{T_{dec}}{T_0}\right)\left(\frac{g_{*dec}}{43/4}\right)^{1/3}\left(
\frac{11}{4}\right)^{1/3},
\end{equation}
where the two factors on the right-hand side include the heating from the annihilation of all particles heavier than electrons, and the
heating due to electron-positron annihilation, respectively.

Assembling all of these factors and using the measured values for $t_{eq}$ and $T_0$, the free-streaming length can be written as
\begin{equation}
\lambda_{FS} = 0.2 ~{\rm pc}~ \left(\frac{5 m_p}{m_\zeta}\right)\left(\frac{m_{\chi}}{T_{dec}}\right)~ g_{*NR}^{1/6}~ g_{*dec}^{-1/3}
\left(1 - 0.04~\ln\left[\left(\frac{5 m_p}{m_\zeta}\right)\left(\frac{m_{\chi}}{T_{dec}}\right)~g_{*NR}^{1/12}
~g_{*dec}^{-1/3} \right]\right).
\end{equation}

A reasonable order-of-magnitude upper bound on $\lambda_{FS}$ to allow structure formation is $\lambda_{FS} \la$ 100 kpc.
Using this limit and appropriate values for the degrees of freedom, we obtain:
\begin{equation}
\label{FSlimit}
\frac{T_{dec}}{m_{\chi}} > 2 \times 10^{-7}\left(\frac{5 m_p}{m_\zeta}\right).
\end{equation}
Roughly speaking, values of ${T_{dec}}/{m_{\chi}}$ no more than an order of magnitude above this bound will
produce $\zeta$ that behaves as
warm dark matter, while larger values of ${T_{dec}}/{m_{\chi}}$ will yield
cold dark matter. 

Clearly, there is a broad range between the limits given in Eqs. (6) and (\ref{FSlimit}). Substituting Eq. (4) into those inequalities gives a range for $M_{GUT}/c_5$

\begin{equation}
2 \times 10^{16} \; {\rm GeV} \left( \frac{m_{\zeta}}{5 m_p} \right)^{1/2} \left( \frac{m_{\chi}}{3 \times
10^6 ~{\rm GeV}} \right)^{1/2} < \frac{M_{GUT}}{c_5}
< 8 \times 10^{18} \; {\rm GeV} \left( \frac{m_{\zeta}}{5 m_p} \right) 
\left( \frac{m_{\chi}}{3 \times 10^6~ {\rm GeV}} \right)^{1/2}.
\end{equation}

We have assumed that a negligible number of $\zeta^c$ are generated by scattering of thermal particles after reheating. In particular, we require that
$n_{\zeta^c}/s < \epsilon_X = (5 m_p/m_{\zeta}) \epsilon_B$.  This gives the following upper limit on the reheat temperature:

\begin{equation}
\left( \frac{n_{\zeta^c}}{s} \right)_{T_{rh}} 
\sim \frac{c_5^{\;2} T_{rh}^3}{M_{GUT}^2} \; \frac{M_{P\ell}}{\sqrt{g_{*rh}} T_{rh}^2}
\; \frac{1}{g_{*rh}} < (5 m_p/m_{\zeta}) \epsilon_B
\; \Longrightarrow \; T_{rh} < \frac{(M_{GUT}/c_5)^2}{3 \times 10^{25} {\rm GeV}}
\left( \frac{5 m_p}{m_{\zeta}} \right).
\end{equation}

All of the bounds we have derived in this section are based on the assumption that there is a strong asymmetry in the dark matter, i.e. $n_{\overline{DM}} \ll n_{DM}$. We regard this as an interesting limit. Moreover, in models where the dark matter asymmetry is generated by sphalerons of a new non-abelian interaction, assuming that limit typically allows a prediction  for the mass of the dark matter particle, as noted in \cite{ADM-nonSMsph}.
But it is interesting also to consider the more general case, where there is a weaker asymmetry in dark matter. This would allow smaller masses for the light stable dark matter particle $\zeta$. It would also allow $M_*$ (and $m_{\chi}$) to be larger, as less efficient pre-annihilation is allowed. It would also relax the upper bound on 
$T_{dec}/m_{\chi}$. We leave a detailed analysis of the more general case to future work
\cite{BCS}.

\section{An $SU(7)$ example}

We now present a simple $SU(7)$ grand unified model that illustrates the ideas discussed above. As noted in the Introduction, grand unification based on large groups has several features that make it a natural context for models of dark matter. (1) The large unified fermion multiplets typically contain SM singlets that could be dark matter candidates, so that such models would {\it unify} dark matter with quarks and leptons. (2) The symmetry that stabilizes the dark matter can arise as an accidental global symmetry in a manner analogous to the way $U(1)_{B-L}$ arises as an accidental symmetry in minimal $SU(5)$. And (3), the large unified groups (if they have rank  $\geq 6$) contain extra non-abelian subgroups that can play the role of $G_*$ in co-generating and pre-annihilating dark matter. That is, the $G_*$ interactions can be unified with the Standard Model interactions. All of these features are realized in the $SU(7)$ model we present in this section.

The group $SU(7)$ contains the subgroups $SU(7) \supset SU(5) \times SU(2)_* \times U(1)_{T_7} \supset G_{SM} \times SU(2)_* \times U(1)_{T_7}$, where $G_{SM}$ is the Standard Model gauge group, and where the generator $T_7$ of the extra abelian factor 
can be written (in the fundamental representation of $SU(7)$) as $T_7 = 
diag (\frac{2}{7}, \frac{2}{7}, \frac{2}{7}, \frac{2}{7}, \frac{2}{7}, -\frac{5}{7}, - \frac{5}{7})$. We assume that $SU(7)$ breaks to $G_{SM} \times SU(2)_* \times U(1)_{T_7}$ at superlarge scales.

The fermions of one family consist of ${\bf 21} + 3 \times \overline{{\bf 7}}$, which we denote 
$\psi^{[AB]}$, $\psi_A$, $\psi^{(k)}_A, k = 1,2$. The indices $A, B, ... $ etc. are $SU(7)$ indices, while $k$ is just a label distinguishing two of the anti-fundamental fermion multiplets. This is the smallest possible anomaly-free set that gives one SM family. In addition, each family is assumed to have three $SU(7)$-singlet fermions that will be denoted simply $\psi^{(m)}$, where $m = 1,2,3$ is a label distinguishing them. 
Besides the Higgs fields that break $SU(7)$ at superlarge scales, there are Higgs multiplets $h^{[ABC]}$, $h'_A$, and $\Omega^{(k)}_A$, $k = 1,2$. These will break the electroweak $SU(2)_L$ and the $SU(2)_*$ and give mass to the fermions, as will be seen. (Here too the index $k$ is just a label distinguishing the two $\Omega_A$ multiplets.)  

We shall use lower-case Greek letters for indices of the $SU(5)$ subgroup of $SU(7)$, so $\alpha, \beta, \gamma, ... = 1, ..., 5$. We shall use capital Latin letters from the middle of the alphabet for indices of the $SU(2)_*$ subgroup of $SU(7)$, so $I, J, ... = 6,7$. We suppress all family indices. The fermion multiplets decompose under $SU(5) \times SU(2)_*$ as shown in Table I. In that table, the $SU(7)$ fermion multiplets of one family are listed across the top row. In the column under each $SU(7)$ multiplet are given the $SU(5) \times SU(2)_*$ multiplets contained in it. The superscripts $0, \pm 1, \pm 2$ on the $SU(5) \times SU(2)_*$ multiplets are the values of the global charge $X$, which will be discussed shortly. The three rows denote three types of fermion: ``SM" stands for Standard Model fermions, which have $X= 0$; ``VL" stands for vectorlike fermions, which have $X = \pm 1$; and ``DM" stands for dark matter fermions, which have $X = \pm 2$.

\noindent {\large\bf Table I:} The decomposition under $SU(5) \times SU(2)_*$ of the fermion multiplets of a family. The superscripts $0$, $\pm 1$, $\pm 2$ are the values of $X$.

\vspace{0.2cm}
\[
\begin{tabular}{|l|l|l|l|l|}
\hline & $\psi^{[AB]} = {\bf 21}$ & $\psi_A = \overline{{\bf 7}}$ & $\psi^{(k)}_A = 2 \times \overline{{\bf 7}}$ & $\psi^{(m)} = 3 \times {\bf 1}$ \\
\hline 
SM & $\psi^{\alpha \beta}$ & $\psi_{\alpha}$ & &  \\
& $= (10, 1)^0$ & $=(\overline{5}, 1)^0$ & & \\  
\hline
VL & $\psi^{\alpha I}$ & $\psi_I$ & $\psi^{(k)}_{\alpha}$ &  \\
& $= (5,2)^{-1}$ & $=(1,2)^{+1}$ & $= 2\times (\overline{5}, 1)^{+1}$ & \\
\hline
DM & $\psi^{[IJ]}$ & & $\psi^{(k)}_I$ & $\psi^{(m)}$ \\
& $= (1,1)^{-2}$ & & $=2 \times (1,2)^{+2}$ & $=3 \times (1,1)^{-2}$ \\
\hline  
\end{tabular}
\]

\noindent Note that, as is required for co-generation of dark matter and ordinary matter by $SU(2)_*$ sphalerons, there are dark matter particles that transform under $SU(2)_*$
(namely those in $(1,2)^{+2}$)
as well as quarks and leptons that do so (namely those in $(5,2)^{-1}$). This is not special to $SU(7)$, but would be a generic feature of models in which $SU(2)_*$ is 
unified with $G_{SM}$ in a simple group.
 
The Higgs multiplets are assumed to have the vacuum expectation values (VEVs) 
$\langle h^{267} \rangle = v_u \sim M_W$, $\langle h'_2 \rangle = v_d \sim M_W$,
and $\langle \Omega^{(k)}_6 \rangle \sim \langle \Omega^{(k)}_7 \rangle \sim M_*$. (The indices of the $SU(2)_L$ electroweak subgroup of $SU(5)$ are $\alpha = 1,2$, while the color indices are $\alpha = 3,4,5$.)
The VEV $v_u$ will give mass to the up quarks; the VEV $v_d$ will give mass to the down quarks and charged leptons; and the VEVs of $\Omega^{(k)}_I $ will give mass to the vectorlike fermions and to the dark matter, as well as breaking $SU(2)_*$. To align these VEVs there must be a coupling connecting these Higgs multiplets. A coupling sufficient to do this is
$h^{[ABC]} h'_A \Omega^{(1)}_B \Omega^{(2)}_C$.

The Yukawa couplings are as follows

\begin{equation}
{\cal L}_{Yukawa} \supset (\psi^{[AB} \psi^{CD} ) h^{EFG]}, \;\; (\psi^{[AB]} \psi_A)
h'_B, \;\; (\psi^{[AB]} \psi^{(k)}_A ) \Omega^{(k')}_B, \;\; 
(\psi^{(k)}_A \psi^{(m)}) (\Omega^{(k')})^{*A}.
\end{equation}

\noindent Assumed {\it not} to be present are couplings similar to those in 
Eq. (16), except with $h'_A \leftrightarrow \Omega^{(k)}_A$. It is easy to check that there is an accidental $U(1) \times U(1)$ global symmetry consistent with all the Yukawa couplings in Eq. (16) and the Higgs self-coupling $h^{[ABC]} h'_A \Omega^{(1)}_B \Omega^{(2)}_C$. We shall be interested in a subgroup of this, which we will call $U(1)_R \times K$, where $K$ is a $Z_2$ symmetry. The $R$ charges and $K$ parities of the various fermion and Higgs multiplets defined so far are given by

\begin{equation}
R^K(\psi^{[AB]}, \psi_A, \psi^{(k)}_A, \psi^{(m)}, h^{[ABC]}, h'_A, \Omega^{(k)}_A)
= (-4^-, 2^+, 9^+, -14^-, 8^+, 2^-, -5^-).
\end{equation}

\noindent When the $\Omega^{(k)}_I = (1,2)$ get non-zero VEVs of order $M_*$, the group $SU(2)_*$ is spontaneously broken, and the group $U(1)_{T_7} \times U(1)_R$ is broken down to a global
symmetry that we shall call $U(1)_X$, whose generator is

\begin{equation}
X= T_7 + \frac{1}{7}R.
\end{equation}

\noindent One easily sees from Eqs. (17) and (18) and the definition $T_7 \cong diag (\frac{2}{7}, \frac{2}{7}, \frac{2}{7}, \frac{2}{7}, \frac{2}{7}, -\frac{5}{7}, - \frac{5}{7})$ that the $SU(5) \times SU(2)_*$ multiplets in Table I have the $X$ values shown there.
One also sees that the Higgs VEVs $\langle h^{267} \rangle, \langle h'_2 \rangle$ 
and $\langle \Omega^{(k)}_6 \rangle \sim \langle \Omega^{(k)}_7 \rangle$ do not break $U(1)_X$.

\noindent The masses of the Standard Model up quarks come from the coupling
$(\psi^{[\alpha \beta} \psi^{\gamma \delta}) \langle h^{267]} \rangle$. Those of the Standard Model down quarks and charged leptons come from $(\psi^{[\alpha 2]} \psi_\alpha) \langle h'_2 \rangle$, with $\alpha$ being a color index (3,4,5) for the down quarks and a weak index (1) for the charged leptons. (Of course, in a realistic model of quark and lepton masses there must always be other terms to avoid the ``bad" unification relation that the charged leptons and down quarks have equal mass at the GUT scale. But there are several ways of adding such terms, and this does not affect anything else we say here.)
The heavy vectorlike fermions  $\psi^{[AI]} = (5,2)^{-1}$ and $\psi^{(k)}_A = 2 \times 
(5,1)^{+1}$ (see the row VL in Table I)) obtain $O(M_*)$ masses from $(\psi^{[AI]} \psi^{(k)}_A) \Omega^{(k')}_I \rangle$. The Yukawa couplings so far mentioned leave the vectorlike fermions $\psi_I = (1,2)^{+1}$ massless. (The coupling $(\psi^{[2I} \psi_I)
\langle h'_2 \rangle$ mixes them with the heavy neutrinos, but leaves the resulting mixture massless.) The masslessness of these fermions would not be a phenomenological problem, but they can be given Dirac masses by introducing some gauge singlet fermions
into the model.

The heavy dark matter particles get mass of $O(M_*)$ from 
$(\psi^{[IJ]} \psi^{(k)}_I) \langle \Omega^{(k')}_J \rangle$ and
$(\psi^{(k)}_I \psi^{(m)}) \langle (\Omega^{(k')})^{*I} \rangle$. The $\psi^{(k)}_I$
(of which there are four per family, as $k = 1,2$ and $I = 6,7$) are what we called $\chi$ in earlier sections. The $\psi^{[IJ]}$ (of which there is one per family) and the
$\psi^{(m)}$ (of which there are three per family) are what we called $\chi^c$. 

In order for the pre-annihilation mechanism discussed in previous sections to work, the $\chi$ and $\chi^c$ must be able to decay to light stable dark matter particles through dim-5 operators that are suppressed by $O(1/M_{GUT})$ or $O(1/M_{P\ell})$. One such operator is $(\psi^{IJ} \zeta) h^*_{[2IJ]} h'^{*2}/M_{GUT}$. Since $\psi^{IJ}$ is one of the $\chi^c$, as noted in the previous paragraph, and $h^{[2IJ]}$ and $h'_2$ are just the weak Higgs doublets $h_u$ and $h_d$ whose VEVs do the electroweak breaking, we may write this operator as $(\chi^c \zeta) h^* h'^*/M_{GUT}$. The $SU(7)$-singlet fermion field $\zeta$ can be assigned $R^K= 14^+$ (implying by Eq. (18) that it has $X = 2$ as the other dark matter particles $\psi^{(k)}_I$ do), in which case this dim-5 operator conserves $R$, $K$ and $X$. 
In order to give a Dirac mass to $\zeta$, let us introduce an SU(7)-singlet fermion $\zeta^c$ that has $R^K = -14^+$ and therefore $X= -2$. (Notice that $K$ parity prevents
the $\zeta$ from having a mass term with $\psi^{(m)}$, and $\zeta^c$ from mixing with  $\psi^{(m)}$.)  

The Dirac fermion made up of $\zeta$ and $\zeta^c$ must have mass of order 1 GeV. The smallness of this mass can be explained in a way that is technically natural \cite{naturalness} by assuming it arises from the dynamical breaking of a symmetry. Let us suppose, therefore that there a $Z_2$ symmetry, which we will call $P$, under which $\zeta^c$ is odd, while all the other fields defined up to this point are even. (That is, all the fermions in Table I, the fermion $\zeta$, and the Higgs fields $h^{[ABC]}, h'_A, \Omega^{(k)}_A$ are even under $P$.)
The $P$ prevents an explicit (or bare) mass term for $\zeta, \zeta^c$. But such a mass can arise dynamically as follows. Suppose that there are terms
$\zeta^c \zeta S$, $\overline{\Psi} \Psi S^*$, and $M_{GUT}^2 S^* S$, where $S$ is a singlet scalar and $\overline{\Psi}$ and $\Psi$ are fermions that transform under a force that confines at a scale $\Lambda$ and acquire a condensate $\langle \overline{\Psi} \Psi \rangle \sim \Lambda^3$. Assume that $S$ and $\Psi^c$ are odd under $P$ while $\Psi$ is even, so as to allow these couplings.
Then $m_{\zeta} \sim \langle S \rangle \sim \Lambda^3/M_{GUT}^2$, which can be of order 1 GeV if the confinement scale $\Lambda \sim 10^{11}$ GeV. 

The mechanism just described for making the mass of $\zeta$ small may seem somewhat contrived, but it is ``natural" in the technical sense \cite{naturalness}, {\it i.e.} in the sense that no coefficient of a renormalizable term in the Lagrangian has to be ``fine-tuned" to be very small. Perhaps a somewhat more elegant way could be found to
make $m_{\zeta}$ small. But this example does serve to illustrate the general fact that the mass of a fermion can be much smaller than the breaking scale of the symmetry that ``protects" it (here the symmetry $P$), without violating the condition of ``technical naturalness". Another example of this (which in fact is closely analogous to the mechanism just described) is the type II see-saw mechanism for neutrino masses, where neutrino masses are protected by the electroweak $SU(2)_L$, but the neutrino masses that result from the breaking of $SU(2)_L$ at the scale
Fermi scale $v$ are only of order $v^2/M_{GUT} \ll v$. 

With all the particles of the model having the $R$ charges and $K$ and $P$ parities given above, it is easy to see that the lowest dimension operator that connects the light dark matter particles $\zeta$ and $\zeta^c$ to the other particles of the low energy theory (i.e. those with mass less than or of order $M_*$) is the dim-5 operator $(\psi^{IJ} \zeta) h^*_{2IJ} h'^{*2}$, which is suppressed by a large ($O(M_{P\ell})$ or $O(M_{GUT})$) mass. That is important, as seen in the previous sections, as otherwise there would be a danger that interactions of $\zeta$ and $\zeta^c$ with other matter would produce a thermal abundance of them and and lead to a large symmetric component of dark matter.

One can see that the quantum number $X$ serves to stabilize not only the dark matter, but also the heavy vectorlike $5 + \overline{5}$ fermions. This is easiest to see in the case (which we shall henceforth assume) that all the vectorlike (VL) fermions, which have odd values of $X$, are heavier than all the DM fermions, which have even values of $X$. Then the lightest particle with odd $X$ will be absolutely stable, and the lightest particle with $X = \pm 2$ will also be stable. That $X$ conservation stabilizes both the dark matter and the vectorlike fermions seems to be a generic feature of the kind of model we are discussing if it is unified in a simple group. If $X$ remained exactly unbroken, there would be residual heavy quarks and leptons from these vectorlike multiplets that would come to dominate the energy density of the universe. It is thus necessary that some small spontaneous breaking of $X$ in the model. It must be large enough for the decay of the heavy vectorlike fermions to be sufficiently fast not to cause cosmological problems, but small enough that (a) it does not cause too rapid decay of the heavy dark matter $\chi, \chi^c$ (i.e. dominating over the decays caused by the dim-5 operator), and (b) $X$-violating processes do not wipe out any asymmetries in dark matter.

Fortunately, there are simple ways that such $X$ violation can be introduced into the model that satisfy these conditions. One way is to suppose that there are $X$-violating VEVs of the Higgs multiplets $h'_A$ and $\Omega^{(k)}_A$ that are of order $\epsilon \ll 1$ times the $X$-conserving VEVs:
$\langle h'_I \rangle \sim \epsilon v_d \sim \epsilon M_W$, and
$\langle \Omega^{(k)}_2 \rangle \sim \epsilon M_*$. Then $\Delta X = \pm 1$ effects are suppressed by $O(\epsilon)$. This would have the following effects. (1) There would be an $O(\epsilon)$ mixing of the SM down quarks $d = \psi ^{[a2]}$ with the heavy vectorlike down quarks $D = \psi^{[aI]}$, so that $D$ could decay into 
$u + W_-$ with an amplitude suppressed by a factor of $\epsilon$. (2) There would be similar mixing of the SM charged leptons $\ell^-$ with the heavy vectorlike charged leptons $L^-$, allowing the latter also to decay weakly with an amplitude suppressed by a factor
of $\epsilon$.
(3) There would be mixing among all the neutral fermions, namely the SM neutrinos ($\nu$), the heavy vectorlike neutrinos ($N$), and the heavy dark matter particles ($\chi$). This would allow weak decay of the heavy dark matter particles into $ \ell^- + W^+$. However, this amplitude would be suppressed by order $\epsilon^2$. The reason is that the diagonalization of the mass matrices of the charged and neutral fermions turns out to give electroweak lepton doublets of the following form:

\begin{equation}
\left( \begin{array}{l} \nu + O(\epsilon) N + O(\epsilon^2) \chi \\
\ell^- + O(\epsilon) L^- \end{array} \right), \;\;\;  
\left( \begin{array}{l} N + O(\epsilon) \nu + O(\epsilon) \chi \\
L^- + O(\epsilon) \ell^- \end{array} \right).
\end{equation}

\noindent The powers of $\epsilon$ are directly traceable to the number of units by which $X$ is violated. Because a decay of a heavy dark matter particle into Standard Model particles violates $X$ by two units, its amplitude is suppressed by $\epsilon^2$.

The condition that the $X$-violating decay of the heavy dark matter be slow compared to that caused by dim-5 operator is that $\epsilon^4 M_* \ll M_*^3/M_{GUT}^2 \Rightarrow 
\epsilon^2 \ll M_*/M_{P \ell}$, which is also the condition that $X$-violating scattering be slow enough not to wipe out the dark matter asymmetries. But that condition still allows the weak decays of the heavy vectorlike quarks and leptons to happen very rapidly; indeed with a lifetime short compared to the Hubble time when $T \sim M_*$. (These issues will be 
analyzed in more detail in a future paper \cite{BCS}.) 

It should be noted that despite the spontaneous violation of $X$, the $\zeta, \zeta^c$ are extremely stable on the time scale of the present age of the universe, since the lowest-dimension operators that couple $\zeta, \zeta^c$ to the light particles of the model have dimension 5 and are therefore suppressed by $O(1/M_{GUT})$ or $O(1/M_{P \ell})$, and also because their decay amplitude to Standard Model particles would be further suppressed by two powers of $\epsilon$. For example, if their decay amplitudes are of order $\epsilon^2/M_{GUT}$, they have a lifetime of $\tau_{\zeta} \sim \left( m_{\zeta}^3 \frac{\epsilon^4}{M_{P\ell}^2}
\right)^{-1}$, which for $m_{\zeta} \sim 5$ GeV and $\epsilon^2 \ll M_*/M_{P \ell}$  gives $\tau_{\zeta} \gg 10^{29}$ yrs. 
 
In the $SU(7)$ model we have presented, both the dark matter particles and heavy vectorlike quarks and leptons carry non-zero $X$. (This is a generic feature of such models when unified in a simple group.) As one can see from Table I, one can write $X = 2N_{DM} + N_{VL}$, where $N_{DM}$ is dark matter number (the number of dark matter particles minus the number of anti-dark matter particles) and $N_{VL}$ is the vectorlike particle number. The contribution of the dark matter fermions and vectorlike fermions to $\Delta X$ in $SU(2)_*$ sphaleron processes cancel, as can be seen from Table I. In other words, $SU(2)_*$ sphaleron processes violate $N_{DM}$ and $N_{VL}$, but do not violate $X$. 
(This is also a generic feature of models of the type we have been discussing when unified in a simple group.) Thus, whatever primordial asymmetry is generated after reheating must have $X = 2N_{DM} + N_{VL} \neq 0$, as otherwise $SU(2)_*$ sphalerons will wipe out the asymmetries in both $N_{DM}$ and $N_{VL}$. (This is analogous to the fact that electroweak sphalerons do not violate $B-L$, so that unless a primordial asymmetry in $B-L$ exists electroweak sphalerons will wipe out both $B$ and $L$.)

\section{Conclusions}

We have presented a scenario in which a new non-abelian gauge interaction, whose group we called $G_*$, is responsible for both co-generating a dark matter asymmetry and pre-annihilating the symmetric component of the dark matter. The co-generation of dark matter with ordinary matter is done by the sphaleron processes of $G_*$, while scattering mediated by the exchange 
of $G_*$ gauge bosons accomplishes the pre-annihilation of dark matter with anti-dark matter, so as to leave almost purely asymmetric dark matter. 

We showed that sufficient pre-annihilation in this scenario requires the scale of $G_*$ breaking $M_*$ to be below about $3 \times 10^6$ GeV.  
The pre-annihilation mechanism requires that there be heavy unstable dark matter particles, an asymmetry in which is initially generated and which later decay to a light stable form of dark matter, which inherits this asymmetry. (These heavy unstable dark matter particles, which we call $\chi$, have mass $m_{\chi} \sim M_*$. The light stable dark matter particles, which we call $\zeta$, should have mass of order 5 GeV to obtain $\Omega_{DM} \sim 5 \Omega_B$.) 
We have shown that these $\chi$ decays must have lifetimes that fall within a certain relatively narrow range.
Shorter lifetimes would cause these decays to regenerate the symmetric component
of dark matter, while longer lifetimes would result in the stable dark matter being too hot. The range for $T_{dec}$ (the temperature when the $\chi$ particles decay) is roughly between $2 \times 10^{-7}$ and $10^{-3}$ times $m_{\chi}$.
We have shown that the $\chi$ particles can have decay lifetimes in the required range, 
if they come from dim-5 operators suppressed by $1/M_{GUT}$ or $1/M_{P \ell}$. 

Finally, we have shown that this scenario can be implemented within the context of grand unified models based on groups of rank $\geq 6$. We presented a fairly simple $SU(7)$ model as an example. This model illustrates how several features of the scenario can arise naturally within grand unification: (1) Standard-Model-singlet fermions that can play the role of the heavy dark matter are contained in the fermion multiplets of the large unified group, thus unifying ordinary quarks and leptons with dark matter; (2) global symmetries that stabilize the dark matter can arise as accidental global symmetries, analogously to how $B-L$ arises as an accidental symmetry in minimal $SU(5)$; and (3) new (non-Standard Model) non-abelian gauge symmetries are left after the breaking of the grand unified group, which can play the role of $G_*$. 

There are several model-building issues that deserve further study. One has to do with the breaking of the global symmetry $X$ that stabilizes dark matter. We have found that in fully gauge unified models that realize the scenario we discuss here, the global symmetry $X$ not only stabilizes dark matter but also stabilizes certain heavy vectorlike fermions that generically exist in such models. $X$ must therefore be slightly broken in such a way that the heavy vectorlike particles decay rapidly, while keeping the dark matter sufficiently stable. We showed one way to do this, but it would interesting to see if there are other better ways. Similarly, the way that the light dark matter particles $\zeta$ were given small mass in a ``technically natural" way could possibly be improved upon.

\section*{Acknowledgements} S.M.B. was
supported in part by the Department of Energy (DE-FG02-12ER41808).  R.J.S. was supported in part by the Department of Energy
(DE-SC0011981).

\end{document}